\documentclass[aps,reprint,twocolumn,amsmath,amssymb]{revtex4-1}

\usepackage{hyperref}
\usepackage{graphicx}
\usepackage{graphics}
\usepackage{amsmath}
\usepackage{comment}

\begin{document}

\title{Compact cold atom gravimeter for field applications}

\author{Yannick Bidel}
\email{yannick.bidel@onera.fr}
\author{Olivier Carraz}
\altaffiliation{Present address: European Space Agency - ESTEC Future Missions
Division (EOP-SF) P.O. Box 299, 2200 AG Noordwijk, The Netherlands}
\author{Ren\'{e}e Charri\`{e}re}
\altaffiliation{Present address: Laboratoire Hubert Curien, UMR CNRS 5516,
B\^{a}timent F 18, Rue du Professeur Beno\^{\i}t Lauras, 42000 Saint-Etienne }
\author{Malo Cadoret}
\altaffiliation{Present address: Laboratoire Commun de M\'{e}trologie LNE-CNAM,
61 rue du Landy, 93210 La Plaine Saint Denis, France}
\author{Nassim Zahzam}
\author{Alexandre Bresson}
\affiliation{ONERA, BP 80100, 91123 Palaiseau Cedex, France}

\begin{abstract}

We present a cold atom gravimeter dedicated to field applications. Despite the
compactness of our gravimeter, we obtain performances (sensitivity $42
\,\mu\text{Gal}/Hz^{1/2}$, accuracy $25\,\mu\text{Gal}$) close to the best
gravimeters. We report gravity measurements in an elevator which led us to the
determination of the Earth's gravity gradient with a precision of
$4\,\text{E}$. These measurements in a non-laboratory environment demonstrate
that our technology of gravimeter is enough compact, reliable and robust for
field applications. Finally, we report gravity measurements in a moving
elevator which open the way to absolute gravity measurements in an aircraft or
a boat.
\end{abstract}

\maketitle


Cold atom interferometer is a promising technology to obtain a highly sensitive
and accurate absolute gravimeter. Laboratory instruments \cite{gravi peter,
best sens muller, gravi syrte} have already reached the performances of the
best classical absolute gravimeters \cite{FG5} with a sensitivity of $\sim
10\,\mu\text{Gal}/\text{Hz}^{1/2}$ ($1\,\mu\text{Gal}=10^{-8}\,
\text{m}/\text{s}^2$) and an accuracy of $5\,\mu\text{Gal}$. Moreover, compared
to classical absolute gravimeters, atom gravimeters can achieve higher
repetition rate \cite{high repetion rate acc} and do not have movable
mechanical parts. These qualities make cold atom gravimeters more adapted to
onboard applications like gravity measurements in a boat or in a plane. Cold
atom gravimeters could thus be very useful in geophysics \cite{geophysics} or
navigation \cite{navigation}. In this context, cold atom sensors start to be
tested on mobile platforms. An atom accelerometer has been operated in a
$0\,\text{g}$ plane \cite{ice nature}. An atom gradiometer has also been tested
in a slow moving truck \cite{gradio camion}. In this article, we present a
compact cold atom gravimeter dedicated to field applications. First, we
describe our apparatus and the technologies that we use to have a compact and
reliable instrument. Then, we present the performances of the gravimeter in a
laboratory environment. Finally, we report gravity measurements in a static and
in a moving elevator.


The principle of our cold atom gravimeter is well described in the literature
\cite{gravi peter} and we summarize in this letter only the basic elements. In
an atom gravimeter, the test mass is a gas of cold atoms which is obtained by
laser cooling and trapping techniques \cite{metcalf}. This cloud of cold atoms
is released from the trap and its acceleration is measured by an atom
interferometry technique. We use a Mach-Zehnder type atom interferometer
consisting in a sequence of three equally spaced Raman laser pulses which drive
stimulated Raman transitions between two stable states of the atoms. In the
end, the proportion of atoms in the two stable states depends sinusoidally on
the phase of the interferometer $\varphi$ which is proportional to the
acceleration $g$ of the atoms along the Raman laser direction of propagation:
\begin{equation}
\varphi=k_{\text{eff}}\, g\, T^2\, ,
\end{equation}
where $k_{\text{eff}}\simeq 4\pi/\lambda$ is the effective wave vector
associated to the Raman transition, $\lambda$ is the laser wavelength and $T$
is the time between the Raman laser pulses.


The description of our gravimeter setup is the following. The cold atoms are
produced and fall in a vacuum chamber made of glass connected to a titanium
part to which are connected a 3 l/s ion pump, getters and rubidium dispensers.
This vacuum chamber is inside a magnetic shield consisting of 4 layers of
mu-metal. The falling distance of the atoms is equal to 6 cm. The sensor head
containing the vacuum chamber, the magnetic shield, the magnetic coils and the
optics for shaping the laser beams and collecting the fluorescence has a height
of 40 cm and a diameter of 33 cm. The gravimeter is placed onto a passive
vibration isolation table (Minus-K). The laser system for addressing $^{87}$Rb
atoms is similar to the one described in reference \cite{laser onera}.
Basically, a distributed feedback (DFB) laser diode at 1.5 $\mu$m is amplified
in a 5 W erbium doped fiber amplifier (EDFA) and then frequency doubled in a
periodically poled lithium niobate (PPLN) crystal. A power of 1 W at 780 nm is
available. The frequency of the laser is controlled thanks to a beatnote with a
reference laser locked on a Rubidium transition. The Raman laser and the
repumper are generated with a fiber phase modulator at 1.5 $\mu$m which
generates side bands at 7 GHz. All the electronics and the optics of the
gravimeter fit in one 19'' rack (0.6 x 0.7 x 1.9 m).


The experimental sequence of the gravimeter consists in the following. First,
$^{87}$Rb atoms are loaded from a background vapor in a 3D magneto-optical
trap. The atoms are then further cooled down in an optical molasses to a
temperature of $1.8\;\mu \text{K}$. Then, the atoms are selected in the state
$F=1, m_F=0$ thanks to a microwave selection. After 10 ms of free fall, we
apply the atom interferometer sequence consisting in three Raman laser pulses
of duration 10, 20 and 10 $\mu$s. The Raman laser pulses couple the state $F=1,
m_F=0$ to the state $F=2, m_F=0$. The time between the Raman pulses is equal to
$T=48\,\text{ms}$. During the interferometer sequence, a vertical uniform
magnetic field of $28\,\text{mG}$ is applied. A radio frequency chirp of
$\alpha/2\pi \sim\, 25.1 \text{MHz}/\text{s}$ is also applied to the Raman
frequency in order to compensate the time-dependant Doppler shift induced by
gravity. Finally, the proportion of atoms in the state F=2 and F=1 is measured
by collecting the fluorescence of the atoms illuminated with three pulses of a
vertical retro-reflected beam of durations of 2, 0.1, and 2 ms. The first and
the last pulses resonant with the $F = 2 \rightarrow  F' = 3$ transition give a
fluorescence signal proportional to the number of atoms in the state F=2 and
the middle pulse resonant with $F = 1 \rightarrow F' = 2$ transition transfers
the atoms from the state F=1 to the state F=2. A rms noise of 0.2\% on the
measured proportion of atoms is obtained with this detection scheme limited by
the frequency noise of the laser. The repetition rate of the experimental
sequence is equal to 4 Hz. The measurement of the proportion of atoms $P$ in
the state $F=2$ versus the radio frequency chirp $\alpha$ leads to interference
fringes given by the formula:
\begin{equation}
 P = P_m -
\frac{C}{2}\cos\left((k_{\text{eff}}\,g-\alpha)T^2\right)\, ,
\end{equation}
where $P_m$ is the mean proportion of atoms in the state $F=2$, $C$ is the
contrast which is equal in our case to $C=0.36$.

The protocol of the gravity measurements is the following. The gravity is
measured by acquiring $P$ from each side of the central fringe i.e. for $\alpha
\simeq k_{\text{eff}}\,g\pm \pi/2T^2$. The sign of $\alpha$ and thus the sign
of $k_{\text{eff}}$ is also changed every two drops in order to eliminate
systematic effects which change of sign with $k_\text{eff}$. In order to follow
slow variations of gravity, the central value of $\alpha$ is also numerically
locked to the central fringe. For each atom drop, the gravity is determined
with the last 4 measurements using the following relations :
\begin{eqnarray}
\alpha_n&=& s \left(\alpha^0_n+(-1)^n \frac{\pi}{2T^2}\right)\nonumber\\
g_n&=&\sum_{i=0}^3 \frac{\alpha^{0}_{n-i}}{4 |k_{\text{eff}} |}
-\frac{1}{{|k_{\text{eff}}|T^2}} \arcsin\left(\sum_{i=0}^{3}\frac{(-1)^{n-i}P_{n-i}}{2C}\right)\nonumber\\
\alpha^0_{n+1}&=&\alpha^0_n- G (\alpha^0_n-|k_{\text{eff}}|g_n)
\end{eqnarray}
where $\alpha_n$ is the radio frequency chirp applied at the n-th drop of the
atoms, $\alpha^0_n$ is the value of the central fringe used at the n-th drop of
the atoms, $s=\pm1$ is the sign of radiofrequency chirp which changes every two
drops, $P_n$ is the proportion of atoms in the state F=2 measured at the n-th
drop, $g_n$ is the gravity measurement at the n-th drop and $G$ is the gain of
the lock of the central value of $\alpha$.

\begin{figure}[h!]
 \includegraphics[scale=0.95]{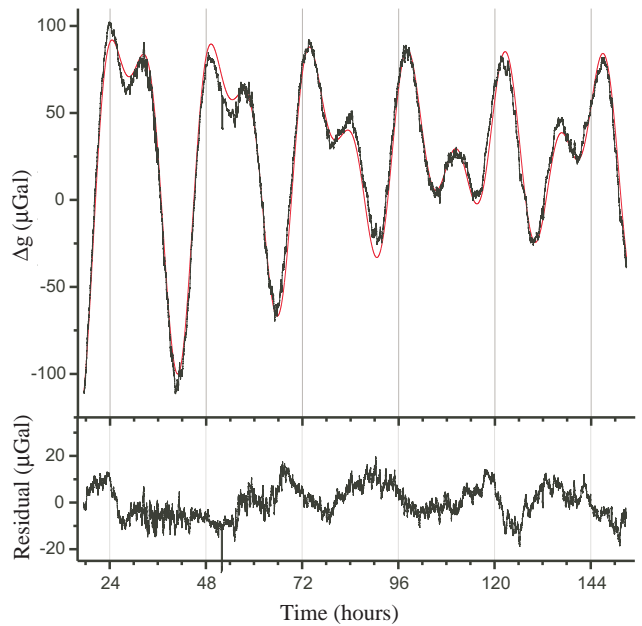}
 \caption{\label{tides} Continuous gravity measurements from 27 May to 2 June 2009. The data are averaged over 15 minutes (3600 atom drops).
 Top: gravity measurements uncorrected from tides with the tide model in red solid line.
 Bottom: residual between the gravity measurements and the tide model.}
\end{figure}
\begin{figure}[h!]
 \includegraphics[scale=0.95]{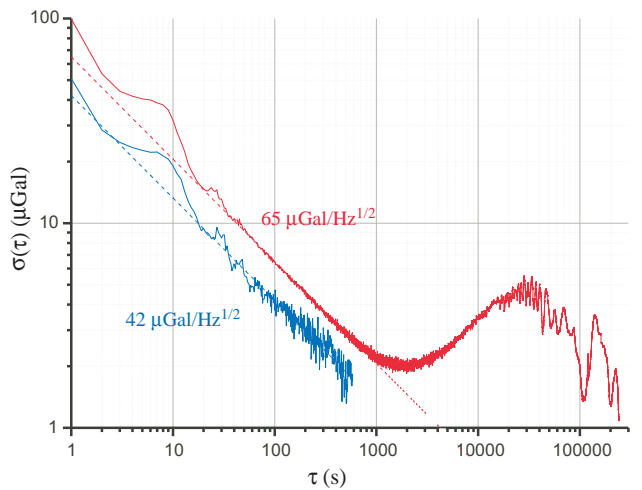}
 \caption{\label{Allan} Allan standard deviation of the gravity measurements.
 The top red line corresponds to the Allan standard deviation of data taken during five days.
 The bottom blue line corresponds to data taken during one night when the vibration level is lower.}
\end{figure}

The gravimeter was tested in our laboratory by acquiring continuously gravity
during five days. The measurements averaged over 15 minutes are shown on Fig.
\ref{tides}. A good agreement is obtained with our tide model \cite{tide model}
with a rms difference of $7\, \mu\text{Gal}$. The Allan standard deviation on
the gravity measurements corrected for the tides is shown on Fig. \ref{Allan}.
A short term sensitivity of $65\,\mu\text{Gal}/\text{Hz}^{1/2}$ is obtained
during the five days of gravity measurements. During the night, when the level
of vibration is lower, one gets a better sensitivity of 42 $\mu$Gal/Hz$^{1/2}$.

This difference of sensitivity between night and day indicates that the
sensitivity of the gravimeter is limited by the vibrations. This is confirmed
by our estimation of the other sources of noise. The detection noise limits the
sensitivity at 15 $\mu$Gal/Hz$^{1/2}$. The phase noise of our microwave source
limits the sensitivity at 2 $\mu$Gal/Hz$^{1/2}$. The frequency noise of the
Raman laser \cite{raman laser frequency noise} limits the sensitivity at $\sim
1\, \mu$Gal/Hz$^{1/2}$.

The main systematic effects which limit the accuracy of the gravimeter were
evaluated and are listed in Table \ref{table}. Our method of generating the
Raman laser by modulation induces a systematic error on the gravity
measurement. This effect was studied in detail in reference \cite{raie
parasite}. In our case, one obtains an uncertainty of 8 $\mu$Gal. The
systematic effects caused by the inhomogeneity of the magnetic field
\cite{gravi peter} and the first order light shift  \cite{gravi peter} change
of sign with $k_{\text{eff}}$. These effects cancel therefore with our protocol
of measurement consisting in alternating the sign of $k_{\text{eff}}$. The
residue of these effects is estimated to be below 1 $\mu$Gal and is negligible
compared to the other systematic effects. The second order light shift \cite{2
photons,2 photons syrte} has been calibrated by measuring gravity versus the
power of the Raman laser. Our uncertainty on the calibration is equal to 2
$\mu$Gal. The Coriolis effect gives an error equal to $2\, v_t\,\Omega$ where
$\Omega$ is the rotation rate of the earth projected in the horizontal plane
and $v_t$ is the transverse velocity of the atoms perpendicular to the Earth
rotation vector. The uncertainty on the transverse velocity of the atoms
detected is estimated in our case at 2 mm/s leading to an uncertainty on $g$
equal to 19 $\mu$Gal. The wavefront curvature of the Raman laser caused by
imperfect optics is causing an error equal to $\sigma_v^2/R$ \cite{gravi syrte}
where $\sigma_v$ is the rms width of the velocity distribution of the atoms and
R is the radius of curvature of the wavefront. We estimate that our optics
induce a wavefront curvature with a radius $|R|$ around 1.4 km leading to an
uncertainty of 12 $\mu$Gal. The Raman laser is aligned vertically by maximizing
the value of gravity. This procedure leads to an uncertainty of 2 $\mu$Gal. The
uncertainty on our laser wavelength is equal to 2 MHz giving an uncertainty of
5 $\mu$Gal. The quadratic sum of all these contributions gives a total
uncertainty of 25 $\mu$Gal. This accuracy estimation of our gravimeter has been
confirmed by the comparison with a relative gravimeter (Scintrex CG-5)
calibrated with an absolute gravimeter. The relative gravimeter gives a
measurement of gravity equal to $980 883 499 \pm 6\, \mu$Gal. Our atom
gravimeter gives $980 883 165 \pm 25\, \mu$Gal. The difference of height
between the two gravimeters is equal to $1.09\pm 0.03\,\text{m}$ leading to a
correction due to vertical gravity gradient of $347\pm10\,\mu$Gal. Finally, one
obtains a difference between the two measurements equal to $13 \pm 28\, \mu$Gal
in agreement with the error bar.

\begin{table}
\begin{tabular}{|l|c|c|}
  \hline
  Effect & Bias & Uncertainty  \\
  & ($\mu$Gal)&($\mu$Gal)\\
  \hline
  Raman laser generated  by modulation  & -18 & 8\\
  Light shift second order              & 43 &  2\\
  Coriolis effect                       & 0  & 19\\
  Wavefront curvature                   & 0  & 12\\
  Verticality                           & 0  & 2\\
  Laser wavelength                      & 0  & 5\\
  \hline
  Total                                 & 25 & 25 \\
  \hline
  \end{tabular}
\caption{Main systematic effects on the gravity measurements.}\label{table}
\end{table}



\begin{figure}[h!]
 \includegraphics[scale=0.58]{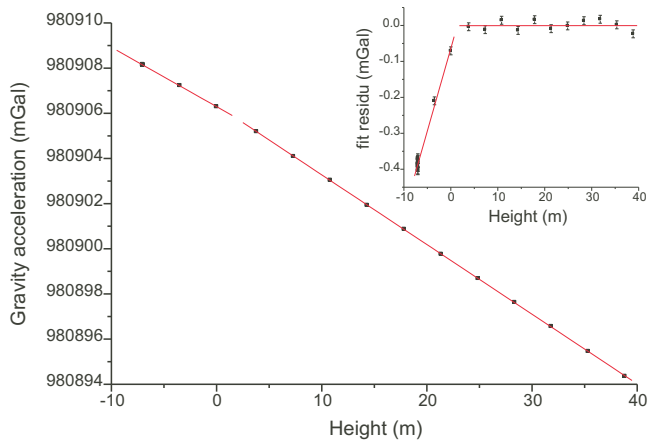}
 \caption{\label{elevator} Gravity measurements versus height in an elevator. The points are the experimental measurements.
 The lines are a linear fit of the measurements overground and underground. The inset in the up right corner is the difference between the measurements
 and the overground linear fit and shows clearly the two slopes which correspond to the gravity gradient overground and underground.}
\end{figure}

The gravimeter was tested in an elevator located in a 14 levels building. A
gravity measurement was done at each level with an acquisition time of 250 s
(1000 drops). The distance between the levels was measured with a laser
distance measurer pointing the top of the elevator cage. At each level, the
verticality of the gravimeter was set thanks to an inclinometer. Between each
gravity measurement at a given level, a gravity measurement at the level -2 was
done in order to check for the repeatability of gravity measurements. The
gravity measurements at the level -2 have a standard deviation of 11 $\mu$Gal.
The gravity measurements at each level are plotted on the Fig. \ref{elevator}.
One can see that the gravity gradient is different above the floor and under
the floor. Overground, a linear fit of the data gives a gravity gradient equal
to $3086\pm4\,\text{E}$. This value agrees with the mean gravity gradient on
the Earth (free-air anomaly) given in the literature \cite{torge}. Underground,
a linear fit of the data gives a gravity gradient equal to  $2626\pm
16\,\text{E}$. The lower underground gravity gradient is due to the mass of the
soil above the measurement point which gives a correction equal to
$4\pi\rho\,G$ where $\rho$ is the density of the soil.

\begin{figure}[h!]
 \includegraphics[scale=0.56]{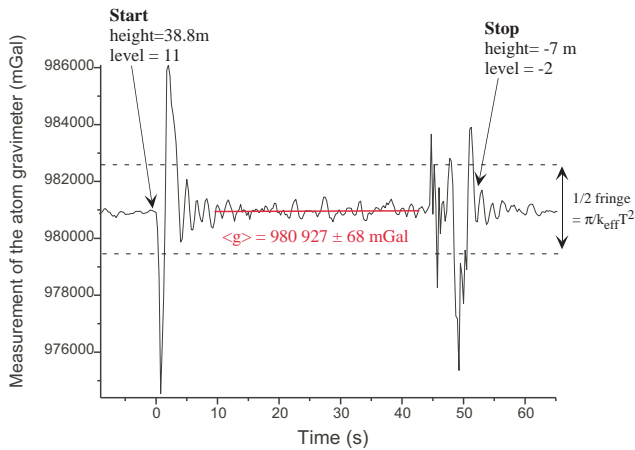}
 \caption{\label{elevator_moving} Gravity measurements in a moving elevator.
 The red line corresponds to the averaged value of the gravity measured during the stabilized part of the descent of the elevator.}
\end{figure}

In order to demonstrate the possibility to measure gravity in a boat or a plane
with an atom gravimeter, we measured gravity while the elevator was moving. To
perform this measurement, our vibration isolation table which can not work
while the elevator is moving was blocked. Thus, the time of the interferometer
$T$ was reduced from 48 ms to 1 ms in order to have variations of acceleration
smaller than one fringe. The measurements of the gravimeter acquired when the
elevator was moving from the level 11 to the level -2 are shown on Fig.
\ref{elevator_moving}. By assuming that the mean acceleration of the elevator
is null during the stabilized part of the descent of the elevator (10 s - 43
s), the measurements in dynamic give a measurement of gravity equal to $980927
\pm 68\,\text{mGal}$ which agrees with the static measurements. The statistical
uncertainty of 68 mGal comes from the acceleration variations of the elevator
and the vibrations.


In conclusion, we demonstrated the possibility to perform quantitative gravity
measurements in a non-laboratory environment with an atom gravimeter. This
demonstration was possible with the development of a compact and robust atom
gravimeter. Despite the fact that we chose a small falling distance in order to
have a compact apparatus, we obtain performances (sensitivity $42
\,\mu\text{Gal}/Hz^{1/2}$ and accuracy $25\,\mu\text{Gal}$) close to the best
gravimeters. Quantitative gravity measurements with a repeatability of
$11\,\mu\text{Gal}$ were performed in an elevator wherein the apparatus is
subject to shocks, vibrations and fluctuations of temperature. These
measurements led us to the determination of the gravity gradient with a
precision of 4 E. We also demonstrated the ability of an atom gravimeter to be
used in a mobile platform by measuring gravity in a moving elevator. Finally,
we point out that technological developments concerning the vibration isolation
system or the association with a classical accelerometer \cite{hybridation, ice
nature} have still to be made in order to have quantitative gravity
measurements in mobile platforms.

We thank the SHOM for their gravity measurements in our laboratory. This work
was supported by the French Defence Agency (DGA).

\end{document}